\documentclass[12pt,preprint]{aastex}
\usepackage{amssymb}

\usepackage {graphicx}

\begin{document}

\title{A note on the discovery of a $2M_\odot$ pulsar}
\author{X. Y. Lai and R. X. Xu}
\affil{School of Physics and State Key Laboratory of Nuclear Physics
and Technology, Peking University, Beijing 100871, China}
\email{laixy@pku.edu.cn}

\begin{abstract}
It is conventionally thought that the state equation of dense matter
softens and thus cannot result in high maximum mass if pulsars are
quark stars, and that a recently discovered $2M_\odot$ pulsar (PSR
J1614-2230) may make quark stars to be unlikely.
However, this standard point of view would be revisited and updated
if quark clustering could occur in cold quark matter because of the
strong coupling between quarks at realistic baryon densities of
compact stars, and it is addressed that the state equation of
clustering quark matter stiffs to support compact stars with maximum
mass $M_{\rm max}>2M_\odot$.
In this brief note, it is demonstrated that large parameter spaces
are allowed for $M_{\rm max}>2M_\odot$ in a Lennard-Jones model of
clustered quark matter, and the newly measured highest mass of PSR
J1614-2230 would be meaningful to constrain the number of quarks
inside single quark-cluster, to be $N_q<\sim 10^3$.

\end{abstract}


\section{Introduction}

Quark stars have been characterized by soft equations of state,
since the asymptotic freedom of quantum chromo-dynamics (QCD) tells
us that as energy scale goes higher, the interaction between quarks
will become weaker. In fact, the simplest and most widely used model
for quark stars, the MIT bag model, treats the quarks inside a quark
star as relativistic and weak-interaction particles which are
confined inside the star by an additional pressure denoted by the
bag constant~\citep[e.g.,][]{Alcock1986}.

However, in cold quark matter at realistic baryon densities of
compact stars (with an average value of $\sim 2-10\rho_0$), the
energy scale is far from the region where the asymptotic freedom
approximation could apply, so the the ground state of realistic
quark matter might not be that of Fermi gas~\citep[see a discussion
given by][]{xu08,xu10}.
The interaction between quarks inside a quark star could make quarks
to condensate in position space to form quark clusters~\citep{xu03},
and at low enough temperature the quark-clusters could crystallize
even to form a solid state of quark stars.
Solid quark stars still cannot be ruled out in both astrophysics and
particle physics~\citep[e.g.,][]{Horvath05,Owen05,mrs07}.

It is really a difficult task to obtain a realistic state equation
of cold quark matter at a few nuclear densities, because of (i) the
non-perturbative effect of the strong interaction between quarks at
low energy scale and (ii) the many-body problem of vast assemblies
of interacting particles. However, it is still meaningful for us to
consider some phenomenological models to explore the properties of
quarks at the low energy scale.
In the earlier work, we tried two models; one is the polytropic
quark star model~\citep{lx09a} which establishes a general framework
for modeling quark stars, and the other one is the Lennard-Jones
quark matter model~\citep{lx09b} which introduces a specific kind of
interactions in quark stars.
Both of the models are much different from the conventional ones
(e.g., MIT bag model). In the former case the quark-clusters are
non-relativistic particles, whereas in the the latter case quarks
are relativistic particles.
Consequently, the equations of state in our two phenomenological
models could be stiffer than that in conventional quark star models,
and then lead to larger maximum masses for quark stars.
Under some reasonable parameters, the maximum mass could be higher
than 2 $M_\odot$.
There could be some other models for the clustered cold quark
matter. \cite{nx2010} adopted a two-Gaussian component soft-core
potential and also found the parameter space in which the maximum
mass could be higher than 2 $M_\odot$.

Recently, radio observations of a binary millisecond pulsar PSR
J1614-2230, which show a strong Shapiro delay signature, imply that
the pulsar mass is 1.97$\pm$0.04 $M_\odot$~\citep{nature2010}.
Although this high mass could rule out conventional quark star
models (whose equations of state are soft) and normal neutron stars
with hyperon or boson condensation cores, the solid quark star model
still cannot be ruled out, because a solid quark star could reach
such high mass without suffering the gravitational instability.
Certainly, besides equation of state, the highest mass would also be
meaningful for researches of $\gamma$-ray bursts (GRBs) and
gravitational waves~\citep{Ozel10} since GRB X-ray Flares may
originate from massive millisecond pulsars produced by compact star
mergers~\citep{Dai06}.

In this note, to constrain the parameters in solid quark stars by
the newly discovered high mass pulsar, we take the Lennard-Jones
model to describe the state of could quark matter in quark
stars~\citep{lx09b}, and present the parameter space which can be
allowed by pulsars with mass higher than $2M_\odot$. We find that if
the number of quarks in one quark-cluster $N_q$ satisfies $N_q<\sim
10^3$, then there is enough parameter space for the existence of
quark stars with masses to be higher than $2M_\odot$.
We also find that the results are consistent with the constraint
imposed by non-atomic spectrum of pulsars.

\section{Constraint on the number of quarks $N_q$ in one quark-cluster}

Quark clustering could occur in cold quark matter because of the
strong coupling between quarks at realistic baryon densities of
compact stars.
The number of quarks in one quark-cluster, $N_q$, is an important
parameter, because it is closely relevant to the strong interaction
between quarks.
To derive the properties of cold quark matter from QCD calculations
is very difficult; however, on the other hand, the astrophysical
observations of pulsar-like compact stars provide us effective tools
to give constrains on the phenomenological models of cold quark
matter.
The constrains by non-atomic spectrum of X-ray observations of
pulsar-like compact stars and by the $2M_\odot$ PSR J1614-2230 can
give us consistent results on the allowed range of $N_q$.

\subsection{Constraint by the non-atomic spectrum}

Strange quark matter is composed of up, down and strange quarks, as
well as electrons to maintain the charge-neutrality.
In MIT bag model, the number of electrons per baryon $N_e/A$ is
found for different strange quarks mass $m_s$ and coupling constant
$\alpha_s$~\citep{FJ1984}.
In their results, when $\alpha_s=0.3$, $N_e/A$ is less than
$10^{-4}$; a larger $\alpha_s$ means a smaller $N_e/A$ at fixed
$m_s$, because the interaction between quarks will lead to more
strange quarks and consequently less electrons.
In our model, we also consider the strong interaction between quarks
as well as between quark-clusters, and consequently the required
number of electrons per baryon to guarantee the neutrality should be
also be very small.
Although at the present stage we have not got the exactly value for
the number density of electrons,  it is reasonable to assume that
$N_e/A$ is less than $10^{-4}$.

Making an analogy between quark-clusters to nuclei, the non-atomic
spectrum of pulsar-like compact stars can give us some implies about
the positive electric charge of an quark-cluster.
The $K_{\alpha}$ line is the strongest line among all the emission
lines of an atom, whose energy is written as
\begin{equation}E_n \simeq -10.2~Z^2~~ {\rm eV},\end{equation}
where $Z$ is the number of positive charge of the nucleus.
Similarly, taking $Z$ as the number of positive charge of one
quark-cluster, from above equation we can get the energy needed for
quark-clusters to emit $K_{\alpha}$ line.
The temperature of a quark star is about 100$\sim$1000 ev, then
$Z<\sim 10$ should be satisfied, otherwise there could be
$K_{\alpha}$ line which is thermally produced.
Consequently, if $N_e/A\sim 10^{-4}$ for each quark-cluster (note
that the baryon number of one quark is 1/3), then $N_q<10^5$ is
required.

\subsection{Constraint by the $2M_\odot$ PSR J1614-2230}

In the Lennard-Jones quark matter model~\citep{lx09b}, the
interaction potential $u$ between two quark-clusters as the function
of their distance $r$ is assumed to be described by the
Lennard-Jones potential~\citep{LJ1924}
\begin{equation}u(r)=4U_0[(\frac{r_0}{r})^{12}-(\frac{r_0}{r})^6],\label{1}\end{equation}
where $U_0$ is the depth of the potential and $r_0$ can be
considered as the range of interaction.
It is worth noting that the property of short-distance repulsion and
long-distance attraction presented by Lennard-Jones potential is
also a characteristic of the interaction between nuclei.
Although the form of interaction between quark-clusters is difficult
for us to derive due to the non-perturbative effect of QCD, we could
adopt the potential in Eq.(\ref{1}) because of its general features.
Like the classical solid, if the inter-cluster potential is deep
enough to trap the clusters in the potential wells, the quark matter
would crystallize and form solid quark stars.

Under such potential, we can get the equation of state, including
the contribution of the lattice vibration inside solid quark stars,
and then derive the mass-radius curves by integrate numerically from
the center to the surface of the star~\citep{lx09b}. In addition,
because of the strong interaction, the surface density $\rho_s$
should be non-zero. The maximum mass of quark stars depends on
parameters $U_0$, $r_0$, $\rho_s$ and the number of quarks inside
one quark-cluster $N_q$.

Given the density of quark matter $\rho$ and the mass of each
individual quark, from Heisenberg's uncertainty relation we can
approximate the kinetic energy of one cluster as $E_{\rm k}\sim 1\
{\rm MeV}(\frac{\rho}{\rho_0})^{\frac{2}{3}}(\frac{N_
q}{10})^{-\frac{5}{3}}$,
where $\rho_0$ is the nuclear matter density.
To get the quark-clusters trapped in the potential wells to form
lattice structure, $U_0$ should be larger than the kinetic energy of
quarks.
Because of the strong interaction between quarks, we adopt $U_0$=10
MeV and 200 MeV to do the calculations.
The surface density $r_0$ should be between 1 to 3 $\rho_0$, to
ensure quark-deconfinement without exceeding the average density for
a typical pulsar. We choose $\rho_s=2\rho_0$ in the calculations.
The minor chance of $\rho_s$ would not chace the results
qualitatively.
In addition, we note that for a given $\rho_s$, we can get $r_0$ at
the surface where the pressure is zero, so there are only three
independent parameters, $U_0$, $\rho_s$ and $N_q$, which determine
the maximum mass of quark stars.

We show relation between the maximum mass of quark stars ($M_{\rm
max}$) and the depth of potential ($U_0$) when $\rho_s=2\rho_0$, for
some different cases of $N_q$, in Fig.1.
%%%%%%%%%%%%%%%%%%%%%%%%%%%%%%%%%%%%%%%%%
\begin{figure}[h]
\begin{center}
  \includegraphics[width=3 in]{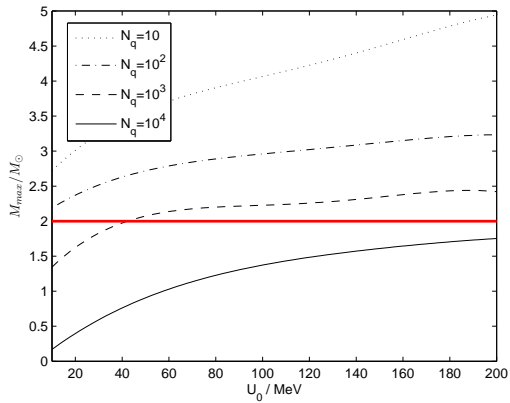}

\end{center}
\caption{%
The dependence of maximum mass $M_{\rm max}$ on $U_0$ (depth of
potential well), for some different cases of $N_q$ (number of quarks
inside one quark-cluster), in Lennard-Jones cold quark matter model.
The surface density $\rho_s$ is chosen to be 2 times of $\rho_0$
(the nuclear matter density). If $N_q<\sim 10^3$, there is enough
parameter space for the existence of quark stars with mass larger
than 2 $M_\odot$.
\label{R}}
\end{figure}
%%%%%%%%%%%%%%%%%%%%%%%%%%%%%%%%%%%%%%%%%%
%
We can see that if $N_q<\sim 10^3$, there is enough parameter space
for the existence of quark stars with mass larger than 2 $M_\odot$.
The case $N_q>10^4$ should be ruled out by the discovery of PSR
J1614-2230.
This constrain of $N_q$ by the maximum mass of pulsars is consistent
with that by the non-atomic spectrum of pulsars ($N_q<10^5$).

Fig.1 also shows that $M_{\rm max}$ is insensitive to $U_0$.
This is understandable, because the repulsive core of the
inter-cluster potential reacts in most part of the dense cold quark
matter inside a quark star, and $U_0$ only reacts near the star's
surface where the density is low enough for one cluster to fell the
depth of the potential well of a nearby cluster.
Furthermore, it could imply that the constraint of $M_{\rm max}$ on
$N_q$ is insensitive to the form of inter-cluster potential, as long
as the potential has a strong repulsive core at short distance.

\section{Conclusions}

The newly discovered high mass pulsar PSR J1614-2230 with mass $\sim
2M_\odot$ still cannot rule out the existence of quark stars,
because quark could be clustered in realistic cold quark matter at
supra-nuclear density and then stiff equations of state are
possible.
We take the Lennard-Jones quark matter model to calculate the
maximum masses of quark stars, finding that if $N_q<\sim 10^3$,
there is enough parameter space for the existence of quark stars
with masses to be higher than 2 $M_\odot$.
Moreover, this constraint on $N_q$ could shows generality for
clustered quark matter, insensitive to the form of inter-cluster
potential.

\end{document}